\documentclass[aps,prb,twocolumn,groupedaddress,showpacs,amsmath,amssymb,amsfonts,floatfix]{revtex4}

\usepackage{graphicx}
\usepackage{bm}

\usepackage{color}
\definecolor{gray}{rgb}{0.6,0.6,0.6}

\begin{document}

\title{Field-Induced Quantum Phases in Frustrated Spin-Dimer Model: \\
A Sign-Problem-Free Quantum Monte Carlo Study}

\author{Kwai-Kong Ng}
\affiliation{Department of Applied Physics, Tunghai University, Taichung 40704, Taiwan}

\author{Min-Fong Yang}
\affiliation{Department of Applied Physics, Tunghai University, Taichung 40704, Taiwan}

\date{\today}

\begin{abstract}
The magnetization process of a frustrated spin-1/2 spin-dimer model on a square lattice is investigated by means of a sign-problem-free quantum Monte Carlo algorithm developed recently. Rich field-induced quantum phases are discovered. We find two spin superfluids satisfying different symmetries upon layer permutation. There exist as well two solid phases with distinct checkerboard patterns. More interestingly, a spin supersolid phase possessing both solid and superfluid long-range orders is observed over a finite regime of magnetic fields. We find that this exotic phase can be stabilized only when small but nonzero spin anisotropy is present. Various field-induced transitions among these phases are explored. Our findings may provide guidance to the search of these interesting quantum phases in real frustrated spin-dimer compounds.
\end{abstract}


\maketitle

\section{Introduction}

Dimerized antiferromagnets are promising candidates for exhibiting fascinating quantum phases of bosonic matters.~\cite{Giamarchi_etal_2008} In the absence of magnetic field, they usually have a gapped spin-singlet ground state, consisting of dimers for closely spaced pairs of spins. Spin-triplet excitations in such systems are well described by interacting bosons. The energy gap to spin-triplet excitations can be closed by applying a large enough magnetic field, such that the lowest triplet excitation begins to condense and the system develops a spin
superfluid (SF) state. This field-induced transition has been experimentally observed in some quantum dimer systems, such as TlCuCl$_3$~\cite{Nikuni00,TlCuCl3_n} and BaCuSi$_2$O$_6$.~\cite{Sebastian06}  Incompressible commensurate crystals of these spin bosons can be stabilized if the effective repulsion between triplets overwhelms their kinetic energy.~\cite{Rice_2002} These crystalline phases break the lattice's translational symmetry and are signaled by the magnetization plateaus.

Besides, an even more exotic spin supersolid (SS) phase, featured by coexistence of superfluid and solid long-range orders, was predicted to exist in some spin-dimer models.~\cite{Ng_Lee_2006PRL,Sengupta_Batista_2007PRL,%
Laflorencie_Mila_2007PRL,Chen_etal_2010,Albuquerque_etal_2011,Murakami_etal_2013}
This phase usually occurs in the vicinity of a magnetization plateau and behaves as an intermediate state between the superfluid and the crystalline ones. The presence of large Ising-like anisotropy and/or magnetic frustration was shown to be essential for the stabilization of this spin SS state. Because strong spin anisotropy is unrealistic for most magnetic systems, possible realizations of this interesting phase would be those materials with frustration. Therefore, accurate theoretical predictions on frustrated dimer magnets are necessary in order to provide guidance to experimental explorations on the spin SS phase and the related phase transitions.

Quantum Monte Carlo (QMC) simulation is often the method of choice in studying quantum many-body systems in two and higher spatial dimensions, since it is numerically exact and intrinsically unbiased.~\cite{QMC_review} However, QMC usually encounters the negative-sign problem in simulating frustrated quantum spin models,~\cite{QMC_review,sign_prob_1} this powerful tool does not seem to be applicable to the systems under consideration. While a generic solution to the sign problem is considered unlikely,~\cite{limit_sign_prob}
one can sometimes rely on specific symmetries of the Hamiltonian to avoid the sign problem for some models.~\cite{sol_1,sol_2,sol_3,Wang_etal_2015,Li_etal_2015,Wei_etal_2016,Li_etal_2016}

Recently, an efficient sign-problem-free QMC method to simulate a broad class of frustrated quantum spin models is proposed.~\cite{Alet_etal_2016,Honecker_etal_2016} Instead of counting on symmetries of the Hamiltonian, a different kind of basis for the Hilbert space is used in these works to avoid the sign problem. This new algorithm is in principle applicable to systems consisting of spin dimers located on bipartite Bravais lattices. Employing the basis of spin eigenstates of \emph{dimers}, it is demonstrated that frustration can lead to no sign problem once the numbers of occurrences for all types of Monte Carlo processes are even or zero. This even-parity condition can be accomplished if appropriate relations among the competing couplings are satisfied.~\cite{Alet_etal_2016} Due to being free from sign problem, large-scale unbiased simulations become possible by using this spin-dimer-based QMC method, and thus novel quantum phase transitions in such systems can be determined precisely.

In this paper, the dimer-QMC approach mentioned above is employed to uncover the possible field-induced quantum phases in a frustrated spin-dimer model on a square lattice. Our system consists of two layers of $S=1/2$ spins interacting antiferromagnetically with a strong interlayer coupling $D$. Here we consider the strongly frustrating case. That is, the intralayer and the frustrating interlayer couplings have the same strength $J$, except a small spin anisotropy $K$ introduced asymmetrically to them [see Fig.~\ref{fig:phase_dia} and Eq.~\eqref{model}]. This spin-dimer system is found to display rich field-induced quantum phases and interesting phase transitions. They are summarized in Fig.~\ref{fig:phase_dia}, which is the main result of this work.

Because of the bilayer structure, two spin SF phases with different canted spin orderings can appear (see Fig.~\ref{fig:phase_dia}): a symmetric SF (sym-SF) phase which is even under permutation of the layers, and an asymmetric SF (asym-SF) phase which is odd in exchanging the layer indices.~\cite{note_1} The latter usually appears in the small-$J$ region and can be understood as a superfluid of the \emph{singlets} (dimers). By contrast, the sym-SF phase at larger $J$ can be considered as a superfluid of the \emph{triplets}.
Besides, three distinct insulating solid phases are found. The interlayer dimer phase always occurs at weak fields $h$ and for small $J$. Upon increasing $J$, the system behaves as an effective $S=1$ easy-axis $XXZ$ model and shows N\'{e}el ordering in the $z$ direction. Interestingly, for intermediate $J$ and at modest $h$, a checkerboard solid (CBS) phase with checkerboard arrangement of the singlets can be established.
Moreover, an interesting spin SS phase is found to emerge in the vicinity of the CBS phase, and the SF order of this spin SS phase has the asym-SF type. Intricate phase transitions among these phases are explored in our study. We note that the N\'{e}el and the spin SS phase can be stabilized only when a weak but nonzero spin-anisotropic coupling $K$ is present. That is, with the help of frustration, novel phases can emerge in nearly spin-isotropic spin-dimer systems. Therefore, our results shed light on the search of such phases in real frustrated compounds with weak spin anisotropy.

This paper is organized as follows: In Sec.~II, the model and the measured order parameters are introduced. Our numerical results are presented and discussed in Sec.~III. We summarize our work in Sec.~IV.

\section{Model and order parameters}

We consider here a Hamiltonian describing $S=1/2$ spins located on a bilayer square lattice, subject to a strong interlayer exchange $D$, a magnetic field $h$, as well as intralayer and frustrating interlayer couplings,
\begin{eqnarray}\label{model}
&&H_{\rm bilayer} = D\sum_{i} \mathbf{S}_{i,1}\cdot\mathbf{S}_{i,2} %
-h\sum_{i,\alpha} S^z_{i,\alpha}  \nonumber \\
&&\;\; +\sum_{\langle i,j \rangle} \left[J_\parallel^\perp \mathbf{S}^\perp_{i,1} \cdot\mathbf{S}^\perp_{j,1} + J_\parallel^z S^z_{i,1}S^z_{j,1} + 1 \leftrightarrow 2 \right] \nonumber \\
&&\;\; +\sum_{\langle i,j \rangle} \left[J_\times^\perp \mathbf{S}^\perp_{i,1} \cdot\mathbf{S}^\perp_{j,2} + J_\times^z S^z_{i,1}S^z_{j,2} + 1 \leftrightarrow 2 \right]  ,
\end{eqnarray}
where the indices $i$ and $j$ denote rungs of the bilayer lattice and $\alpha=1$, 2 labels two different layers. The summation $\langle i,j \rangle$ runs over pairs of adjacent rungs. $\mathbf{S}^\perp_{i,\alpha}$ represent the vectors formed by two transverse components ($x$ and $y$) of these spins, and $S^z_{i,\alpha}$ denote their $z$ components. Here we consider the strongly frustrating case, in which the intralayer and the frustrating interlayer couplings, $J_\parallel$ and $J_{\times}$, have the same strength, while a small spin anisotropy is introduced asymmetrically to them. To be specific, they are parameterized as follows: $J_\parallel^z=J_\times^z=J$ in the $z$ direction, but $J_\parallel^\perp=J+K$, and $J_\times^\perp=J-K$ in the transverse directions. Both the intralayer and the frustrating interlayer couplings become spin-isotropic and have the same strength $J$, if the spin anisotropy $K$ in the transverse direction is turned off. Here, all the couplings, $D$, $J$, and $K$, are assumed to be positive (antiferromagnetic) and $D\equiv 1$ is set to be the energy unit. In this work, we consider a weak spin anisotropy $K=0.1$.

\begin{figure}
\includegraphics[trim={2.5cm 0cm 0.5cm 0.5cm},clip,width=3.8in]{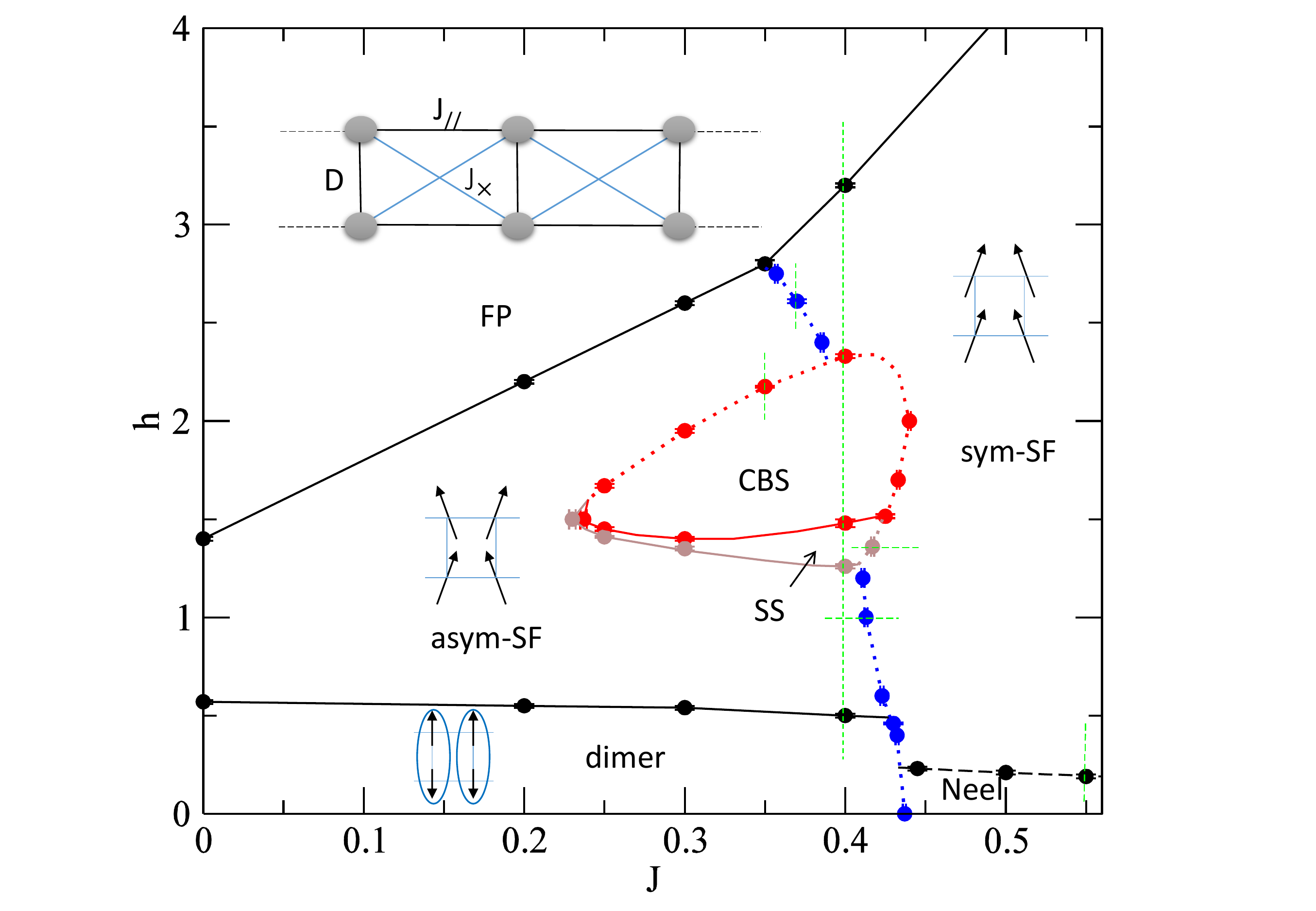}
\caption{(Color online) Ground-state phase diagram for the spin-1/2 bilayer spin model in Eq.~\eqref{model} with the spin anisotropy $K=0.1$, where  $K=J_\parallel^\perp - J_\parallel^z=-(J_\times^\perp - J_\times^z)$ and $J= J_\parallel^z=J_\times^z$. Here the interlayer coupling $D\equiv 1$.
The dotted (solid) lines indicates the first (second) order phase transitions. Six (green) dashed lines correspond to the cuts for which various transitions across different quantum phases are examined below.}
\label{fig:phase_dia}
\end{figure}

We note that the $z$ components of the intralayer and the frustrating interlayer couplings have equal strength in the present model. As shown in Ref.~\onlinecite{Alet_etal_2016}, such a model on a bipartite lattice allows for efficient QMC simulations in a sign-problem-free way, if the canonical basis of \emph{dimers} is employed. Here the eigenstates in this basis are taken to be $|s\rangle = ( |\!\uparrow\downarrow\rangle - |\!\downarrow\uparrow\rangle )/\sqrt{2}$, %
$|t_{+}\rangle = -|\!\uparrow\uparrow\rangle$, %
$|t_{-}\rangle = |\!\downarrow\downarrow\rangle$, and %
$|t_0\rangle = ( |\!\uparrow\downarrow\rangle + |\!\downarrow\uparrow\rangle )/\sqrt{2}$. %
The sign-problem-free condition in the present case amounts to the prohibition against direct transitions between the $|s\rangle$ and $|t_0\rangle$ states on each rung. According to the bond-operator theory,~\cite{bond_op_1,bond_op_2,bond_op_3} quantum transitions among these eigenstates can be described in terms of four types of bosonic operators, in which $s^{(\dagger)}$ and $t_m^{(\dagger)}$ destroy (create) the singlet and the triplet states of the dimer, respectively. Here the label $m=\pm$, 0 refers to the threefold triplet multiplet. The number densities of the singlet and the triplet states are denoted by $n_{s}$ and $n_{t_m}$, respectively. By taking advantage of this dimer basis, we perform the stochastic series expansion (SSE) QMC simulations~\cite{QMC_review} on the model in Eq.~\eqref{model} of sizes $N=L\times L$ rungs ($L=24$ for the main results). Periodic boundary conditions are assumed. Temperature is taken as low as $T=0.01$, so that the ground-state properties are ensured.

Some technical details need to be taken carefully in order to identify the phases correctly. It has been previously pointed out that, at low temperatures and close to quantum phase transitions, the thermalization could be very slow caused by the ineffective mixing of local spin states in this dimer-based approach.~\cite{Alet_etal_2016,Honecker_etal_2016} In some cases the true ground state cannot be easily reached when low-lying metastable states are present. To alleviate this problem, we approach the transition point gradually either by lowering $T$ from higher temperatures or by tuning the coupling $J$ or field $h$ from stable states away from the phase boundaries. Nevertheless, a typical thermalization can take about $~5\times 10^{5}$ Monte Carlo steps (MCs) and even up to a few $10^6$ MCs around the phase boundary of the Neel phase, where strong first-order phase transitions occur.

Many interesting phases can appear in this frustrated dimer system. Some of them can be detected simply by the number densities of the singlet and the triplet states. For example, the dimer phase with the singlet state on each rung should have $n_{s}=1$ and $n_{t_m}=0$ for all $m$. In addition, the fully polarized (FP) phase, in which all spins are polarized along the direction of the external field, will give $n_{t_+}=1$ and other number densities being zero. These characteristics can be rephrased in the spin language. Using the mapping between the spin and the bosonic operators,~\cite{bond_op_1,bond_op_2,bond_op_3} we have %
$S^z_1 + S^z_2 = N_{t_+} - N_{t_-}$ on each rung, where $N_{t_m}$ denotes the local occupation number of the $|t_m\rangle$ state. Therefore, the $z$-component uniform magnetization %
$m^z \equiv (1/N) \sum_i \left\langle S^z_{i,1} + S^z_{i,2} \right\rangle$ has the value of $m^z=0$ ($m^z=1$) in the dimer (FP) phase.

More order parameters are necessary to characterize other phases.
The quantum solid order with checkerboard arrangement of the $|t_+\rangle$ states can be signaled by a finite value of the static structure factor $S({\bf Q}) \equiv (1/N^2)\sum_{i,j} \left\langle (S^z_{i,1} + S^z_{i,2}) (S^z_{j,1} + S^z_{j,2}) \right\rangle e^{i{\bf Q}\cdot({\bf r_i}-{\bf r_j})}$, where ${\bf Q}=(\pi,\pi)$. This can be understood from the spin-boson mapping, $S^z_1 + S^z_2 = N_{t_+} - N_{t_-}$, mentioned in the last paragraph. We note that, in the present study, there are two possible solid orders of checkerboard pattern with nonzero $S({\bf Q})$. One involves the $|t_+\rangle$ and the $|s\rangle$ states, while the other relates to the $|t_+\rangle$ and the $|t_-\rangle$ states. The former case is named as the checkerboard solid (CBS) order and has $n_{t_+}=n_{s}$. To distinguish with the CBS one, we call the latter case with $n_{t_+}=n_{t_-}$ as the N\'{e}el order because of its resemblance to that in an $S=1$ $XXZ$ model with easy-axis spin anisotropy.

As usual, the spin SF order directly links to nonzero spin stiffness (superfluid density) %
$\rho_s = (\langle w_x^2 \rangle + \langle w_y^2 \rangle )/2\beta$, where $w_a$ is the winding number along the $a$ direction. In our study, we find two distinct SF orders with finite $\rho_s$. One has nonzero $n_s$ but $n_{t_0}=0$, while the other has $n_s=0$ but finite $n_{t_0}$. These characteristics can be understood as well in the spin language, as depicted in Fig.~\ref{fig:phase_dia}. By expressing the spin operators in terms of the bosonic operators,~\cite{bond_op_1,bond_op_2,bond_op_3} we have
$S^+_1 - S^+_2 = \sqrt{2} ( s^\dag t_- + t_+^\dag s )$ %
and %
$S^+_1 + S^+_2 = \sqrt{2} ( t_0^\dag t_- - t_+^\dag t_0 )$ on each rung, where $S^+_\alpha = S^x_\alpha + iS^y_\alpha$. These expressions show that the $s$ operator always relates to the asymmetric combination of the spins of two layers, while the $t_0$ operator appears only in the symmetric combination for the in-plane components. Hence the SF state with nonzero $n_s$ corresponds to the ordered phase with nonzero in-plane staggered magnetization of $S^+_1 - S^+_2$, which is odd under layer permutation. We thus call this phase as the asymmetric SF (asym-SF) phase. Alternatively, the SF phase with nonzero $n_{t_0}$ is related to the in-plane staggered magnetization of $S^+_1 + S^+_2$ being even in exchanging the layer indices, and is named as the symmetric SF (sym-SF) phase.

\section{Numerical results and discussions}

\begin{figure}
\includegraphics[trim={5.0cm 0.5cm 5.5cm 0.2cm},clip,width=3.5in]{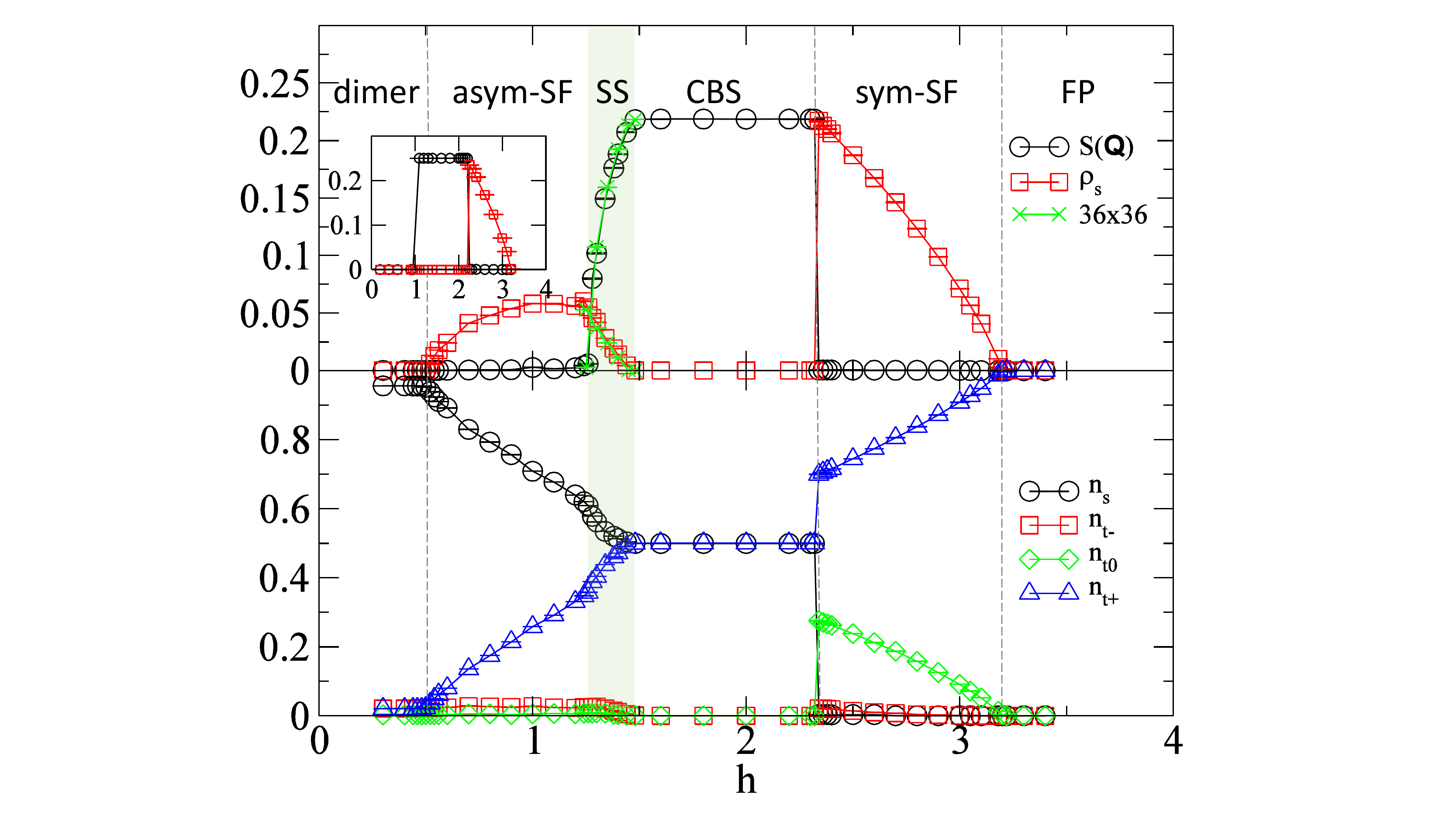}
\includegraphics[trim={5.0cm 0.5cm 5.5cm 0.2cm},clip,width=3.5in]{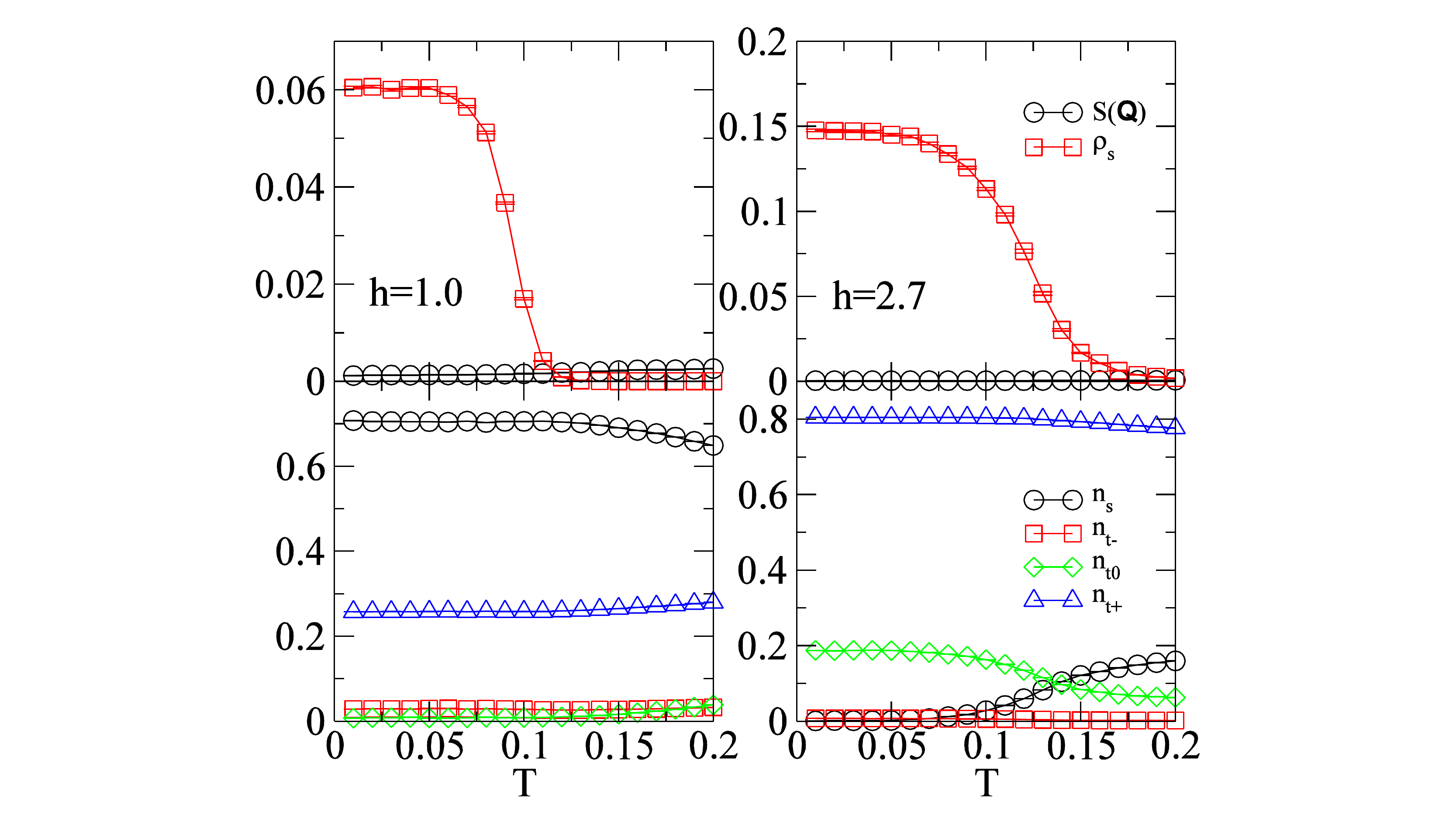}
\caption{(Color online) Upper panels: values of $S({\bf Q})$, $\rho_s$, $n_{s}$, and $n_{t_m}$ ($m=\pm$, 0) for the ground states at $J=0.4$ as functions of the external field $h$ for systems of size $24\times 24$ with $T=0.01$. Result of larger system size $36\times 36$ is also shown for the SS phase to demonstrate that the finite size effect is negligible. The same scan at $J=0.4$ but $K=0$ is plotted in the inset. Lower panels: temperature dependence of various order parameters at $h=1.0$ and $h=2.7$ for systems of size $24\times 24$ with $J=0.4$.}
\label{fig:J=0.4}
\end{figure}

We begin with the $J=0.4$ case, in which most of the aforementioned quantum phases occur upon increasing the external field $h$. Our results are shown in Fig.~\ref{fig:J=0.4}. The zero-field dimer phase will undergo a transition at $h=0.5$ into the asym-SF phase with nonzero values of $\rho_s$ and $n_s$. As $h\geq 1.26$, the system enters the spin SS phase with coexistence of the SF and the solid orders, and then becomes the CBS phase with $S({\bf Q})\neq 0$ and $n_{t_+}=n_{s}=1/2$ for $h\geq 1.48$. Because all the values of $S({\bf Q})$, $\rho_s$, and four number densities are changed continuous at these critical fields, the corresponding transitions are found to be continuous. 

We note that a small but nonzero value of $n_{t_-}$ allows the spin SS phase to be energetically favored against phase separation for the present square-lattice model. In the case of only $n_s$ and $n_{t_+}$ being finite, the system  effectively reduces to a hardcore boson model with only nearest-neighbor interactions, in which no SS phase exists as shown in previous studies.~\cite{28a,28b} Besides, reducing the coupling $J$ will weaken the checkerboard ordering and then leads to the absence of both the CBS and the spin SS phases. This can be understood from the fact that the out-of-plane couplings $J_\parallel^z=J_\times^z=J$ provides the repulsion between the $t_+$ particles and gives a positive contribution to the energy of single-particle excitations out of the CBS state. Therefore, the excitation gap of the incompressible CBS phase can vanish as $J$ decreases such that the CBS state becomes unstable. It results in a closed curve of the CBS phase boundary in the low-$J$ side of the phase diagram, as shown in Fig.~\ref{fig:phase_dia}.

Further increasing $h$, $S({\bf Q})$ will drop to zero abruptly at $h= 2.33$. In the meantime, $\rho_s$ and $n_{t_0}$ will jump to nonzero values. This indicates a first-order melting transition from the CBS phase into the sym-SF phase. Eventually, the system becomes fully polarized with $n_{t_+}=1$ for $h\geq 3.2$. We find that there exists no spin SS phase in the spin-isotropic case with $K=0$ (see inset of Fig.~\ref{fig:J=0.4}). It implies that small but nonzero spin anisotropy could be essential for the stabilization of the spin SS phase.

To ensure that our measurements at $T=0.01$ do provide the ground-state properties, various order parameters as functions of temperature $T$ for two typical cases, $h=1.0$ and $h=2.7$ with $J=0.4$, are shown in the lower panels of Fig.~\ref{fig:J=0.4}. The former (the latter) case gives the asym-SF (the sym-SF) state at zero temperature. It is found that, as reducing temperatures, the systems will eventually transit into the SF phases with nonzero $\rho_s$. In addition, all the measured quantities approach to constants when $T<0.05$. Thus the temperature $T=0.01$ we used is indeed low enough to give the zero-temperature values.

Now we turn our attention to discuss some details on these phase transitions. The instability of the dimer phase is triggered by energy gap closing of the one-triplet excited state. On the other hand, the FP state becomes unstable when the lowest excited state with a single spin flip becomes degenerate with the FP state. Both transitions are thus expected to be of second order and in the universality class of the dilute Bose gas quantum-critical point.~\cite{Sachdev:book} Because the states with a single spin flip are the eigenstates of our model in Eq.~\eqref{model} and their energy eigenvalues can be calculated \emph{exactly}, the exact expression for the saturation field $h_{s}$ can be derived. We find that, when $J\geq 0.35$ ($J\leq 0.35$), $h_s = 2zJ$ [$h_s = 1+ z(J+K)$], caused by the gap closing of the lowest $|t_0\rangle$ ($|s\rangle$) state. Here $z=4$ is the coordination number of square lattices. Therefore, the sym-SF (asym-SF) order will develop below $h_s$. As seen from the phase diagram in Fig.~\ref{fig:phase_dia}, our QMC findings for the phase boundaries of the FP state agree well with analytic predictions. On the other hand, the continuous transition between the asym-SF and the SS states is expected to be of Ising type, because only the discrete translational symmetry is spontaneously broken beyond this critical field. In contrast, the continuous SS-CBS transition should belong to the superfluid-insulator universality class,~\cite{SF-MI} since the $O(2)$ rotational symmetry along the $z$ direction is restored after this transition.

\begin{figure}
\includegraphics[trim={5cm 0.5cm 7cm 0cm},clip,width=3.5in]{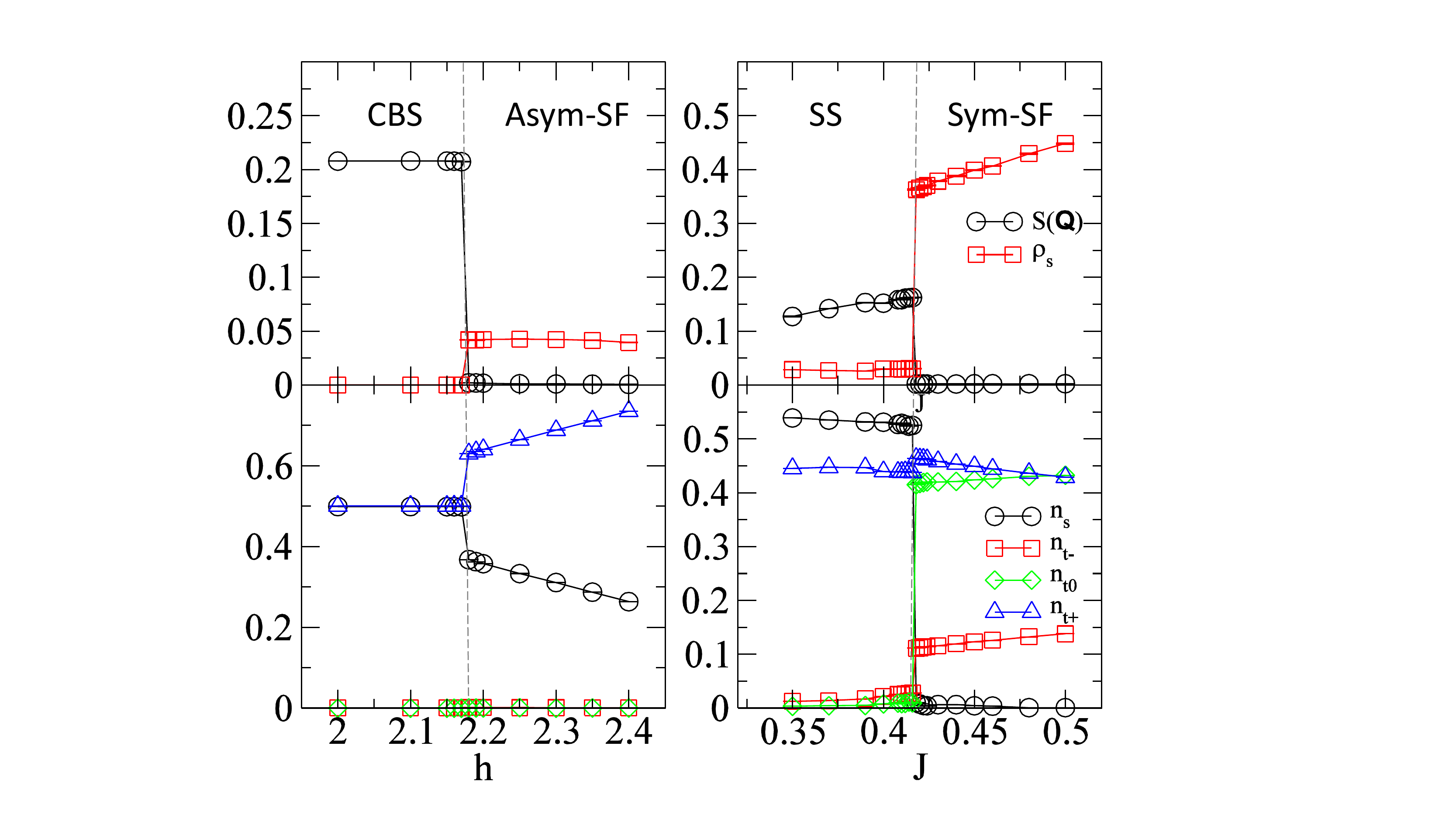}
\caption{(Color online) Left panel: first-order transition from the CBS to the asym-SF phases by increasing $h$ for $J = 0.35$. Right panel: first-order transition from the spin SS to the sym-SF phases by varying $J$ for $h=1.36$.}
\label{fig:other_trans}
\end{figure}

More phase transitions related to these phases are given in order.
Firstly, since the SF and the crystalline states break different symmetries, according to the Ginzburg-Landau-Wilson paradigm, a direct transition between these two phases must be of first order. Otherwise, an intermediate SS phase with coexistence of SF and solid long-range orders will emerge in between. The occurrence of the SS phase between the asym-SF and the CBS phases at modest fields has been established in Fig.~\ref{fig:J=0.4}. For different system parameters, direct first-order transitions from the CBS to the asym-SF phases can also be realized at higher $h$'s. This is illustrated in the left panel of Fig.~\ref{fig:other_trans}, where $\rho_s$ jumps to a nonzero value and $n_s$ drops suddenly below $1/2$ at $h= 2.175$ for $J = 0.35$. The upper lobe of the CBS phase boundary depicted in Fig.~\ref{fig:phase_dia} is partly formed by such first-order transitions.
Secondly, besides continuous transitions to the asym-SF states, the spin SS states can be turned into the sym-SF states as well. However, such transitions are found to be of first order as displayed in the right panel of Fig.~\ref{fig:other_trans}, in which $S({\bf Q})$ drops abruptly to zero associated with a discontinuous change in $\rho_s$ at $J\simeq 0.417$ for a fixed $h=1.36$. The origin of the discontinuous nature of this transition comes from the fact that the SF order ($n_s\neq 0$ but $n_{t_0}=0$) carried by this SS state has distinct character from that of the sym-SF states ($n_{t_0}\neq 0$ but $n_s=0$). Detailed discussions on this point are presented in the next paragraph.
Finally, in the present case, direct transitions from the the dimer to the sym-SF phases are allowed. As seen in the phase diagram in Fig.~\ref{fig:phase_dia}, instead of continuous transitions to the asym-SF states, the transitions to the sym-SF phase are found to be of first order.

\begin{figure}
\includegraphics[trim={5cm 0.5cm 5cm 0.2cm},clip,width=3.5in]{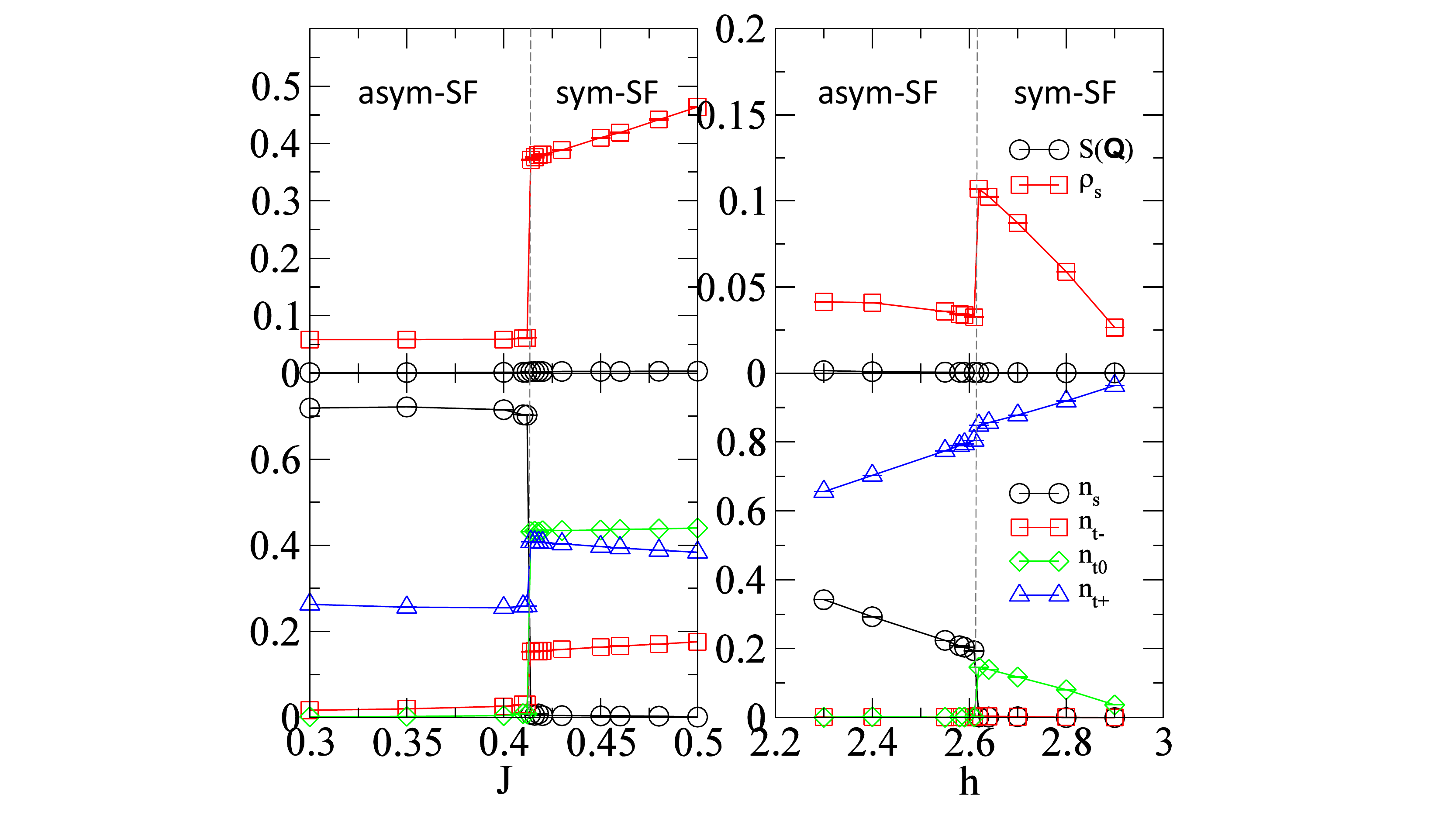}
\caption{(Color online) Phase transitions between two spin SF phases. Left (Right) panel shows the values of the order parameters around the transition point at $J= 0.413$ for $h=1.0$ (at $h=2.61$ for $J=0.37$) located on the lower (upper) phase boundary in Fig.~\ref{fig:phase_dia}.}
\label{fig:two_SF}
\end{figure}

As discussed above, there are two kinds of the spin SF orders, the sym-SF and the asym-SF ones, in the present spin-dimer system. They are respectively related to the in-plane staggered magnetizations of $S^+_1 + S^+_2$ and $S^+_1 - S^+_2$, and thus obey different $Z_2$ symmetry upon layer permutation. A direct transition between these two SF phases has been reported in Ref.~\onlinecite{Murakami_etal_2013}, where a similar frustrated bilayer model was considered. Within their mean-field analysis, this transition is found to be of first order. Thanks to the dimer-QMC method, now we can explore this issue in a numerically exact way. As seen in the phase diagram in Fig.~\ref{fig:phase_dia}, the phase boundaries separating these two SF phases consist of two parts in the present case. Typical behaviors of the order parameters in varying system parameters are illustrated in Fig.~\ref{fig:two_SF}. Transitions at both parts of the phase boundaries are found to be discontinuous, indicated by the discrete jumps in $\rho_s$, $n_{t_0}$, and $n_{s}$. Since these two SF phases satisfy distinct $Z_2$ permutation symmetry, the discontinuity exhibiting at these phase transitions is expected. This explains as well the discontinuous transition between the spin SS and the SF phases with distinct kinds of SF orders, observed in the right panel of Fig.~\ref{fig:other_trans}.

\begin{figure}
\includegraphics[trim={5.0cm 0.5cm 5.5cm 0.2cm},clip,width=3.5in]{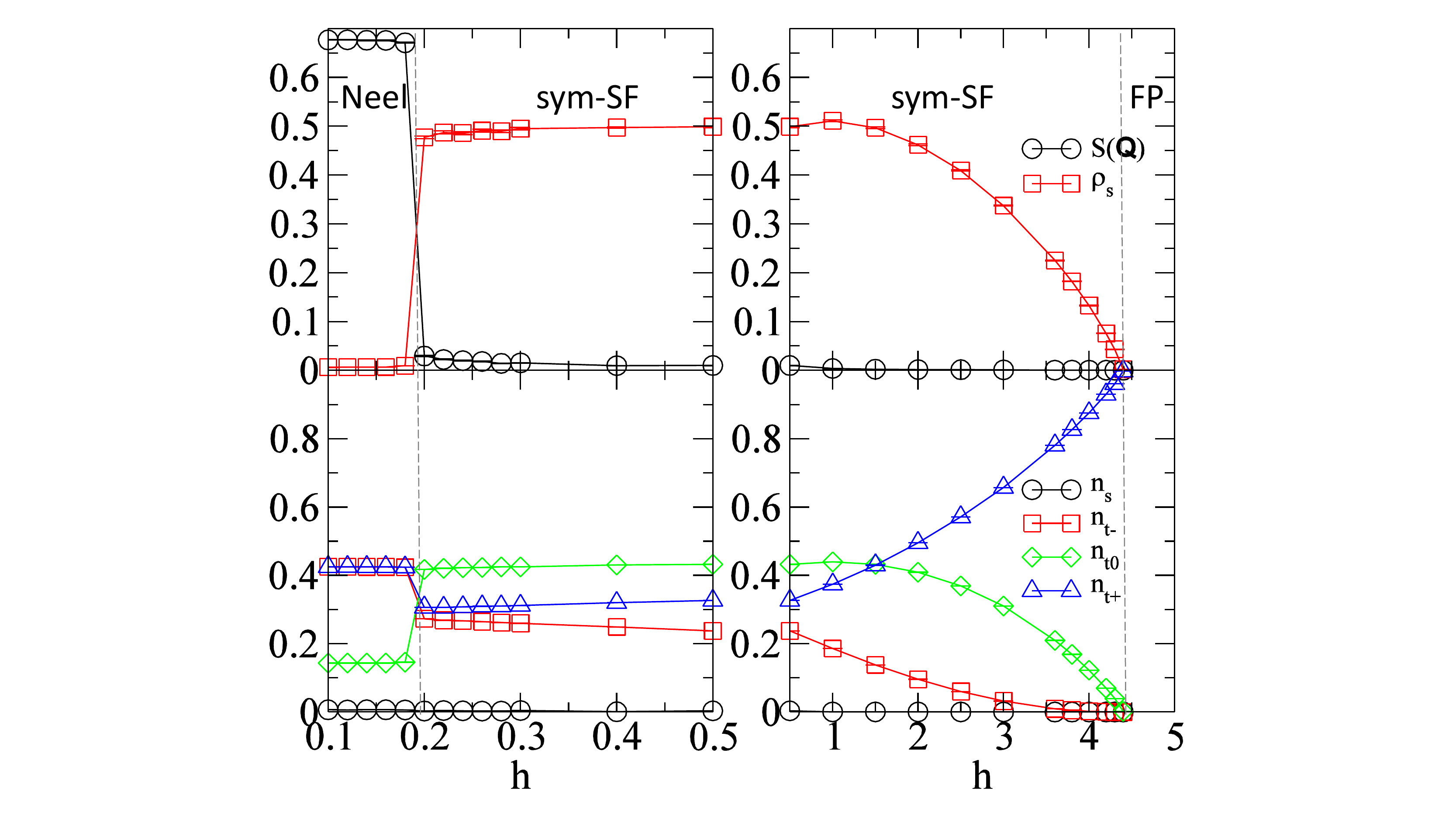}
\caption{(Color online) Values of $S({\bf Q})$, $\rho_s$, $n_{s}$, and $n_{t_m}$ ($m=\pm$, 0) for the ground states at $J=0.55$ as functions of $h$.
}
\label{fig:J=0.55}
\end{figure}

Starting from other zero-field states, different field-induced phases can be discovered. As seen from the phase diagram in Fig.~\ref{fig:phase_dia}, the gapped N\'{e}el state, rather than the dimer one, will become stabilized at zero field when $J\geq 0.437$. This gapped N\'{e}el state has a finite staggered magnetization in the $z$ direction signaled by nonzero static structure factor $S({\bf Q})$, but has no SF order ($\rho_s=0$). In resemblance to the ground state of an $S=1$ easy-axis $XXZ$ model, this state is basically formed by the checkerboard arrangement of the $|t_+\rangle$ and the $|t_-\rangle$ states. According to previous investigations on the easy-axis $XXZ$ model,~\cite{Kohno_Takahashi_1997,Yunoki_2002,Holtschneider_etal_2007}
upon increasing fields, a direct first-order transition to the spin-flop phase (i.e., the sym-SF phase here) is expected. The field-induced transitions for $J=0.55$ is plotted in Fig.~\ref{fig:J=0.55}. When $h\geq 0.19$, a direct first-order transition from the gapped N\'{e}el state with $S({\bf Q})\neq 0$ and $n_{t_+}=n_{t_-}$ to the sym-SF phase with $\rho_s$ and $n_{t_0}\neq 0$ but $n_s=0$ is confirmed, as shown in the left panel of Fig.~\ref{fig:J=0.55}. Eventually, the system becomes fully polarized for $h\geq 4.4$ (see the right panel of Fig.~\ref{fig:J=0.55}). We note that different magnetization process will appear when the spin anisotropy $K$ is absent. In the spin isotropic case with $K=0$, the zero-field state coincides with the \emph{gapless} ground state of an $S=1$ spin-isotropic Heisenberg model. Upon increasing magnetic fields, this gapless state will smoothly evolve into the sym-SF state and will become fully polarized when $h$ is larger than the saturation field.

\section{Summary and conclusions}

Employing the spin-dimer-based QMC simulations proposed in Ref.~\onlinecite{Alet_etal_2016,Honecker_etal_2016}, interesting quantum phases and related field-induced transitions in a strongly frustrated spin-dimer model are revealed. There exist two different solid orders with checkerboard pattern (CBS and N\'{e}el), two distinct SF orders (asym-SF and sym-SF), and one spin SS phase carrying asym-SF orders. Various field-induced transitions among these phases are determined. We note that the gapped N\'{e}el and the spin SS phase are found to be stabilized only when a weak but nonzero spin-anisotropic coupling $K$ is present. Therefore, in combination with the effects of frustration, weak spin anisotropy is enough to give interesting field-induced phases. Our findings may guide the experimental search for spin-dimer compounds exhibiting these novel phases.

\emph{Note added in proof.}--- %
After this work was completed, we notice that a QMC study on the same model was reported in the revised version of Ref.~\onlinecite{Alet_etal_2016}, where only some properties at finite temperatures were discussed.

\begin{acknowledgments}
K.K.N. and M.F.Y. acknowledge the support from the Ministry of Science and
Technology of Taiwan under Grant No. MOST 105-2112-M-029-006 and MOST 105-2112-M-029-005, respectively.
\end{acknowledgments}

\end{document}